\documentclass[conference]{IEEEtran}
\IEEEoverridecommandlockouts
\usepackage{amsmath,amssymb,amsfonts}
\usepackage{algorithmic}
\usepackage{graphicx}
\usepackage{textcomp}
\usepackage{xcolor}
\usepackage[square,numbers,comma]{natbib}
\usepackage{booktabs}
\usepackage{listings}

\def\BibTeX{{\rm B\kern-.05em{\sc i\kern-.025em b}\kern-.08em
    T\kern-.1667em\lower.7ex\hbox{E}\kern-.125emX}}

\usepackage{listings}
\usepackage{xcolor}

\definecolor{codegreen}{rgb}{0,0.6,0}
\definecolor{codegray}{rgb}{0.5,0.5,0.5}
\definecolor{codepurple}{rgb}{0.58,0,0.82}
\definecolor{backcolour}{rgb}{1,1,1}

\lstdefinestyle{mystyle}{
    backgroundcolor=\color{backcolour},   
    commentstyle=\color{codegreen},
    keywordstyle=\color{magenta},
    numberstyle=\tiny\color{codegray},
    stringstyle=\color{codepurple},
    basicstyle=\ttfamily\footnotesize,
    breakatwhitespace=false,         
    breaklines=true,                 
    captionpos=b,                    
    keepspaces=true,                 
    numbers=left,                    
    numbersep=5pt,                  
    showspaces=false,                
    showstringspaces=false,
    showtabs=false,                  
    tabsize=2
}

\begin{document}

\title{On the experiences of adopting automated data validation in an industrial machine learning project \\

\thanks{The research in this paper has been supported by Software Center, the Chalmers Artificial Intelligence Research Centre (CHAIR) and Vinnova.}}

\author{\IEEEauthorblockN{Lucy Ellen Lwakatare\IEEEauthorrefmark{1},
Ellinor R\r{a}nge\IEEEauthorrefmark{2}, Ivica Crnkovic\IEEEauthorrefmark{3} and
Jan Bosch\IEEEauthorrefmark{4}}
\IEEEauthorblockA{\IEEEauthorrefmark{1}\IEEEauthorrefmark{3}\IEEEauthorrefmark{4}Department of Computer Science and Engineering,
Chalmers University of Technology\\
Email: \IEEEauthorrefmark{1}llucy@chalmers.se,
\IEEEauthorrefmark{3}ivica.crnkovic@chalmers.se,
\IEEEauthorrefmark{4}jan.bosch@chalmers.se}
\IEEEauthorblockA{\IEEEauthorrefmark{2}Ericsson\\
Email:\IEEEauthorrefmark{2}ellinor.range@ericsson.com}
Gothenburg, Sweden
}

\maketitle

\begin{abstract}
\textit{Background}: Data errors are a common challenge in machine learning (ML) projects and generally cause significant performance degradation in ML-enabled software systems. To ensure early detection of erroneous data and avoid training ML models using bad data, research and industrial practice suggest incorporating a data validation process and tool in ML system development process. 

\textit{Aim}: The study investigates the adoption of a data validation process and tool in industrial ML projects. The data validation process demands significant engineering resources for tool development and maintenance. Thus, it is important to identify the best practices for their adoption especially by development teams that are in the early phases of deploying ML-enabled software systems.

\textit{Method}: Action research was conducted at a large-software intensive organization in telecommunications, specifically within the analytics R\&D organization for an ML use case of classifying faults from returned hardware telecommunication devices. 

\textit{Results}: Based on the evaluation results and learning from our action research, we identified three best practices, three benefits, and two barriers to adopting the data validation process and tool in ML projects. We also propose a data validation framework (DVF) for systematizing the adoption of a data validation process. 

\textit{Conclusions}: The results show that adopting a data validation process and tool in ML projects is an effective approach of testing ML-enabled software systems. It requires having an overview of the level of data (feature, dataset, cross-dataset, data stream) at which certain data quality tests can be applied.

\end{abstract}


\section{Introduction}

Developing real-world operational machine learning (ML)-enabled software systems require the application of sound software engineering (SE) principles and practices. The SE principles and practices focus on (tool-supported) methods and approaches to ensure the systematic design of robust and reliable software systems. These ensure that the focus of ML system development process goes beyond the consideration of only ML models ~\cite{Amershi2019}. 

ML-enabled software systems rely on high-quality input data to train ML models, which identify useful patterns in data and perform inference on new data using the learned patterns. When erroneous data is fed into ML models, the problems encountered are different than in traditional software. Not only that the output for particular data is wrong, but it can influence the construction of a incorrect or substandard model. Also, even when the output of the model is (sufficiently) correct, its performance will significantly degrade over time.

Data errors are common and can be difficult to detect when developing and operating ML-enabled software systems~\cite{breck2019Google, Schelter2018Amazon, Krishnan2016ActiveClean}. For companies, data errors can result in significant losses in business value. For example, LinkedIn observed financial losses and had to put huge efforts to detect data errors in their job recommendations platform~\cite{Swami2020LinkedIn}. Poor visibility of complex data dependencies, errors in application code, drifts in sensor data, gaps in data due to network connection problems are among the causes of data errors~\cite{schelter2018challenges, Nurmined2019, pizonka2018domain}. Understanding the different types of data errors and their effects on ML projects is important because literature shows that unnecessary data cleaning can be wasteful and harmful to the training of ML models~\cite{qi2018impacts}. 

To handle data errors in ML projects, research and industrial practice suggest integration of data validation tools into the development process of ML-enabled systems\footnote{AI-enabled systems are the systems that include AI components, typically ML components created in a ML training process.} instead of only relying on data scientists to manually check the quality of the data ~\cite{Ehrlinger2019Siemens, breck2019Google,Schelter2018Amazon,hynes2017Google,uberEngineering2020}. Important data quality dimensions of consideration are with respect to \textit{accuracy, completeness, consistency, timeliness}~\cite{Schelter2018Amazon, ehrlinger2019survey}. The data validation tools are particularly useful when dealing continuously with large scale data~\cite{breck2019Google, hynes2017Google, Schelter2018Amazon, Swami2020LinkedIn}. The data validation process is also considered an approach to testing ML-enabled software systems \cite{Braiek2020MLTesting}. 

While several problems in existing data validation tools can be identified, including implementation errors and decoupling from data cleaning capabilities~\cite{ehrlinger2019survey}, much focus is on the implementations of these different tools ~\cite{breck2019Google,hynes2017Google,Schelter2018Amazon,Swami2020LinkedIn}. There is limited reporting on the experiences of adopting the data validation process. The experiences are especially useful for teams that are in the early stages of deploying to production ML-enabled software systems. Adopting the data validation process and tool demands huge engineering resources for development and maintenance~\cite{Swami2020LinkedIn}. Furthermore, there are no well-established guidelines for establishing a data validation process sufficient to guarantee data quality as a whole. In practice, the data validation process is cumbersome, and most engineering teams choose to ignore its tools in their workflows if they do not meet certain requirements~\cite{Swami2020LinkedIn}. For instance, if it introduces additional wait time causing workflows to serve stale data~\cite{Swami2020LinkedIn}.

In this study, we conducted an action research study to investigate the adoption of data validation process and tool for ML projects within analytics R\&D organization at a large telecommunication company. The main research questions (RQs) are:
\begin{itemize}
    \item RQ 1. What are the best practices of adopting a data validation process and tool for an ML project?
    \item RQ 2. What are the benefits of adopting a data validation process and tool for an ML project?
    \item RQ 3. What are the barriers to adopting a data validation process and tool for an ML project?

\end{itemize}

The main contributions of the paper are as twofold. First, the paper presents the experiences of adopting data validation process and tool in an ML project as well as its benefits and barriers at a large software-intensive organization. Second, the paper proposes a framework useful to help systematize the adoption of data validation process and tool in industrial ML projects. 

The rest of the paper is organised as follows. Background and related work is presented in Section \ref{sec:background}. 
In Section \ref{sec:methodology}, action research methodology as applied in the study is discussed. Study results--best practices, benefits and barriers of adopting a data validation process and tool in ML projects--are presented in Section \ref{sec:findings}. The proposed framework is presented in Section \ref{sec:DVF}. The findings in relation to prior literature are discussed in Section \ref{sec:discussion} and concluded in \ref{sec:conclusion}.

\begin{figure*}[ht!]
	\centering
	\includegraphics{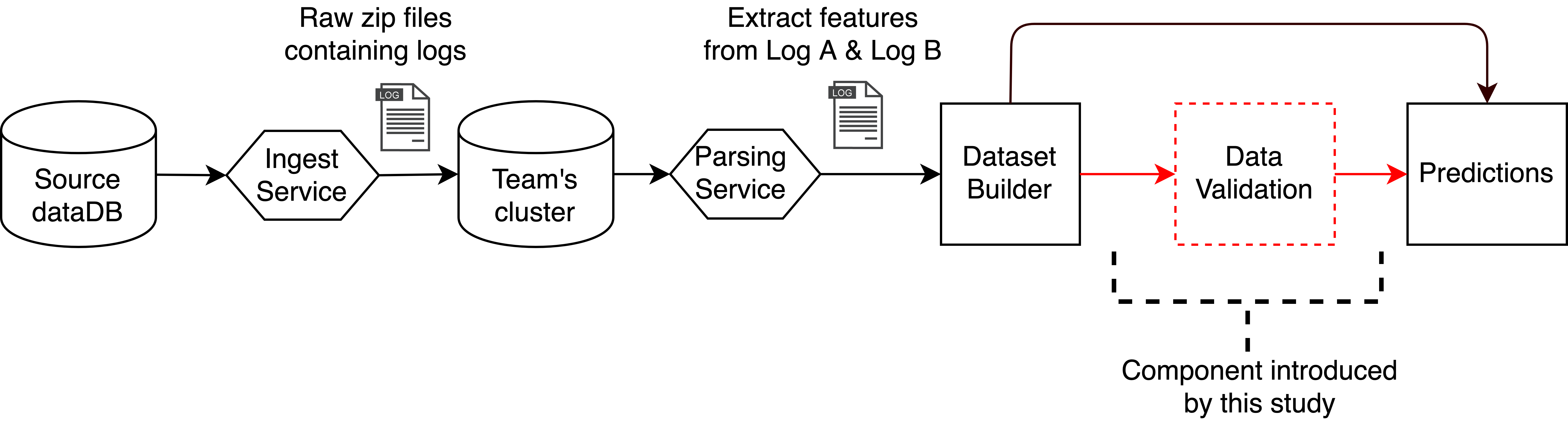}
	\caption{The ML model development process with the extended data validation introduced by this study}
	\label{fig:data-pipeline}
\end{figure*}

\section{Background and related work}
\label{sec:background}
This section presents empirical studies on data validation processes and tools for ML projects. 

\subsection{Data validation process in ML projects}
In most commercial ML systems, deployed ML models are continuously retrained in order to adapt to environmental changes~\cite{breck2019Google, Hazelwood2018Facebook}. When retraining the ML models, new training data collected at inference time can have different distribution due to various reasons, like bugs in application code~\cite{breck2019Google, Sutton2018datadiff}. Differences in data distribution at training and inference, also called training-serving skew, is one form of data errors in ML projects. When the erroneous data is not detected, ML models are retrained on problematic data and can result to performance degradation of an ML system~\cite{breck2019Google, Sutton2018datadiff}. Furthermore, it is rare that training datasets collected from many different sources at different time periods would always have the same exact structure and distribution~\cite{Sutton2018datadiff}.

Data validation in ML projects is the process of ensuring the high quality of data that is fed into the ML algorithm(s). The aim is to continuously check and monitor the data in order to assess its quality and identify underlying issues in data quality~\cite{breck2019Google,Schelter2018Amazon}. Recently, studies have demonstrated that the ML model performance increases when data quality is continuously monitored and corrected according to the data quality measurement results~\cite{Ehrlinger2019Siemens}. 

\subsection{Data validation tools in ML projects}
Ehrlinger et. al.~\cite{ehrlinger2019survey} conducted a state-of-the-art survey of data quality systems (both commercial and open-source), and investigated their measurement and monitoring functionalities in order to determine how data quality is measured and monitored. While their survey did not include other tools identified by this study (discussed in next paragraphs), several limitations are reported, including implementation errors and narrow coverage of data quality metrics for important quality dimensions~\cite{ehrlinger2019survey}. In addition, their study did not report the actual use of the data quality tools in industrial ML projects~\cite{ehrlinger2019survey}. We identify and present studies that discuss the use of data validation tools in industrial ML projects.

A tool called \textit{Data Linter} (adopting the concept of code lint in SE) is used to automatically inspect training data and suggest ways in which features can be transformed into suitable data representation~\cite{hynes2017Google}. The assumption is that data can be valid but not in a representation that the ML model can best learn from, e.g. a timestamp encoded as a string. Three types of data lints that can be detected by the tool are miscoding lints (e.g. number as string), lints for outliers and scaling (e.g. tailed distribution detectors) and packaging error lints (e.g. duplicate values). Technically, the data linter tool inspects training dataset's summary statistics, examines individual examples and names given to the data features. One main limitation of the data linter tool is that it does not allow users to configure and select a set of specific lint detectors to run. As a result, the latter affect tool performance especially for large and medium scale dataset~\cite{hynes2017Google}.The tool does not provide support for data transformation, rather the user has to manually perform the transformations. This is in addition to the lack of proper documentation and discontinued support of the data linter tool \footnote{Data Linter: https://github.com/brain-research/data-linter}. 

\textit{Deequ} is a tool developed by Amazon Research for automating data quality verification. Deequ allows its users to define `unit tests' for data and combines common quality constraints with user-defined validation code~\cite{Schelter2018Amazon}. To perform data validation, the tool relies on declarative user-defined checks on the dataset, for example, isComplete and isUnique checks. The declarative user-defined checks are converted into computations of metrics on data, e.g. different statistical analysis, that can be used to evaluate constraints. After executing data quality verification, the tool reports constraints that succeeded and failed, including information of the computed metric. Although Deequ provides overall data quality report, the tool does not fetch individual records that did not succeed the validations.

At Google, the \textit{TensorFlow data validation} tool~\cite{breck2019Google} is used to validate trillions of training and serving examples per day. To perform data validation, the tool relies on a data schema that is defined and updated, but the first version can also be automatically generated by the tool based on the analysis of sampled data. The data schema is a versioned description of the expected properties of data. Prior to training an ML model, properties of new input data e.g. a dataset from inference service is validated against the data schema in single batch validation to detect data errors within each batch. Any mismatch is flagged as an anomaly that corresponds to a violation of some property specified in the data schema. Validation warnings include suggestions of how to improve the schema. For inter-batch validation to detect data errors between two or more batches, statistical analyses particularly distance metrics such as cosine similarity, are used to quantify the distance between data distributions and detect skew. Similar to Deequ, the tool does not fetch individual failed records. Further, the tool is optimized for Google's TensorFlow ML platform (TFX) that integrates several components needed to define, deploy and monitor ML systems and is based on TensorFlow ML framework~\cite{Baylor2017tfx}.

Finally, Data Sentinel \cite{Swami2020LinkedIn} is a data validation platform developed at LinkedIn. To perform data validation, users use a well-structured configuration file to specify data checks that are desired for specific features. This simplifies the need to write and maintain data checking code. For a given dataset, Data Sentinel computes statistical summaries of the specified features and evaluates the assertions. Eventually, the summaries and validation results are recorded into a dataset profile and validation report.

Overall, studies do not provide experiences of adopting a data validation process and tool by development different teams. The tools presented are also developed by dedicated teams in large companies with several years of experience in deploying to production several ML projects. The few studies that share experiences show slow and poor early adoption with several development iterations \cite{Swami2020LinkedIn}. For companies that are in the early stages of deploying ML components to production and from the embedded domain, learning from these experiences is important to help systematize the adoption with minimum resources. This is because the data validation process and tools consume huge amounts of engineering resources and maintenance~\cite{Swami2020LinkedIn}.

\section{The Action research case study}
\label{sec:methodology}
A participatory action research~\cite{Petersen2014} was used and selected in this study because there was an opportunity for researchers and practitioners \textit{i.e., problem owners} to collaboratively work together to bring about change. Five iterative steps of action research process cycle namely, \textit{(1) diagnosis, (2) action planning (3) action taking, (4) evaluation and (5) specifying learning}~\cite{Petersen2014, mckay2001dual, Staron2020} were applied in the study and are described in detail below. 

\subsection{Action research problem diagnosis and field organization}

\noindent The diagnosis step of action research focuses on identifying, understanding and describing the problem from the industrial context~\cite{Petersen2014, Staron2020}. Data scientists in an R\&D analytics organisation ($\sim$100 persons) at a large software-intensive company within telecommunication domain had problems in the development process of ML-enabled system. In particular, inefficiencies in the development process and infrastructure for ML component of predicting and classifying faults in returned hardware (HW) units caused system crashes and errors that resulted to problems, like giving corrupted files during data ingestion and incorrect data for an ML model. Low data quality and inability to quickly identify data problems and their sources consumed huge development efforts. 

In this case study, we have formed an action-research team to address these challenges and to find appropriate solution. The team included 1) three researchers, one of them actively participating in development and analysis of the results with the practitioners, 2) the core team of three practitioners - two data scientists and a ML engineer who worked in the project on the daily basis, and 3) a larger group of five selected practitioners, data scientists, software developer and a line manager, who participated in tool demonstration and in the evaluation discussions. 

\subsection{Action planning}
\noindent In action planning, different methods for solving the problem are identified, discussed and a choice on how to solve the problem is made~\cite{Petersen2014, Staron2020}. 

In this case study, the data scientists required a systematic way of identifying potential issues in ML training datasets. ML training datasets are used to develop ML models for classifying faults in returned hardware (HW) units e.g., basebands and radios. The HW Return Fault categorization model gives the probability of each class of fault (HW fault, SW fault, No fault) and an indication of which class is predicted. Since the ML component is operational at a screening center, validating ML training datasets against desired properties of data was seen as an appropriate approach to help detect issues arising from the data collection process. ML training dataset is created by primarily extracting data from HW log files and an external tool. The external tool provides log analysis results information on whether the tool has passed or failed assessment. 

All raw data is collected from a central storage (source dataDB). From source dataDB, an ingestion service is used to collect the raw data and store it into the data science team's HDFS based cluster. Data collection to Source dataDB is outside the control of the team. It was agreed that a data validation step and tool would be introduced into the development process. Figure \ref{fig:data-pipeline} shows where data validation is introduced from the original development process of building ML models. An initial plan was to have detailed understanding of the actual and desired characteristics of data while also study and consider to use existing data validation tools~\cite{breck2019Google, Baylor2017tfx,hynes2017Google,Schelter2018Amazon}. 

\subsection{Action taking}

\begin{table}[b]
  \caption{Characteristics of ML training dataset}
  \label{tab:dataset}
  \setlength{\tabcolsep}{0.6\tabcolsep}
  \centering
  \begin{tabular}{*{2}{l}l}
    \toprule
    \textbf{Source} & \textbf{ Extracted Features} & \textbf{Type} \\
    \midrule
    {Log A} &\textit{34 Pattern based}, number of lines matching a pattern & Integer \\
            & \textit{168+ dynamic pattern based}, all instances of a pattern & Integer \\ 
            & \textit{Total lines}, total number of lines in the logs & Integer  \\
            & \textit{Entry rate}, average time between records in log & Float  \\
            & \textit{Duration}, time between the last and first entry in the log & Integer  \\ \hline
    Log B & \textit{16 Pattern based}, number of lines matching a pattern & Integer  \\
            & \textit{Total lines}, number of lines in the log & Integer  \\ \hline
    
    Tool & \textit{Fail/Pass} log analysis result, if it failed & String\\
        & \textit{List of hits}, a list of rules that failed & String \\
        & \textit{Number of hits}, the number of failed rules & Integer \\
    \bottomrule
  \end{tabular}
\end{table}

The chosen method for the solution development is implemented in the action taking step~\cite{Petersen2014, Staron2020}. The explored existing data validation tools~\cite{breck2019Google, Baylor2017tfx,hynes2017Google,Schelter2018Amazon} were not used, but rather a prototype tool was developed. This was due to technological dependencies required by the tools e.g., TensorFlow and Apache Beam for TensorFlow data validation~\cite{breck2019Google} and limited support of the used programming language e.g., Scala in Deequ~\cite{Schelter2018Amazon} among others. However, our prototype tool incorporated the learnings gained from studying the existing data validation tools. It adopts a simplified implementation of \textit{Deequ}~\cite{Schelter2018Amazon} for quality checks and Tensorflow Data Validation~\cite{Baylor2017tfx} for schema validation. 

Data scientists first described desired properties of ML training datasets based on the assumptions of the ML model. In dataset builder, over 200 rules (for external tool) and regex patterns (for logs) are used to create features of ML training dataset. Furthermore, before the training process, some feature transformation is done e.g. one-hot encoding and filling all null values with 0. Apache Spark is used for data transformation and building ML training dataset, which is then stored in an artifact repository. The ML training dataset of approximately size 30 MB on average is stored in batches in the repository. Each batch of ML training datasets includes several hundreds of features and thousands on data instances. Table \ref{tab:dataset} summarises the characteristics of ML training dataset. The properties of data were used to specify and derive data quality checks of the data validation step. 

Our prototype tool uses Pandas\footnote{Pandas Data Analysis: https://pandas.pydata.org/} and Sci-kit learn\footnote{Sci-kit learn: ML algorithms https://scikit-learn.org/}. The user defines validation checks on data, which takes as input a Pandas' DataFrame  and the user can also define specific features to validate a set of constraints (uniqueness, size, completeness, range etc) depending on the preference as shown in Listing 1. Using Pandas, the tool computes statistics on the data (analyser module) and validates (validate module) based on constraints on data (constraints module).

\begin{lstlisting}[language=Python, caption= Simplistic example of data quality test]
from validate import Validation
from metadata import Metadata

# Create dataframe of the first batch
df = Metadata("Data/", "data_set_2.csv").create_dataframe()

# Validate first batch against schema
data_validation = Validation(df)
data_validation.schema_validator("Schema/schema_1.json")

# Run the tests
data_validation.duplicated()
data_validation.outliers()
\end{lstlisting}

The tool presents quality validation as output showing the computed metrics and their results as shown in Listing 2. The user can opt to retrieve and show records containing data errors because a requirement for the data validation prototype tool was to allow users to identify records that violate a given constraint.

\begin{lstlisting}[language=Python, caption = Output example of data quality check]
Schema validation
{'New features': 163, 'Not in Min-Max':1}
Duplicated
{'Dataset': duplicate ratio: 0.3365276211950395'}
Outliers
{'Dataset': outlier ratio: 0.08887636226982337'}
\end{lstlisting}

\subsection{Evaluation}
\label{eval}
At the evaluation step, the effects of action taking are captured using methods such as focus groups, interviews, observations and  questionnaires~\cite{Petersen2014,Staron2020}. To evaluate efficiency of data validation prototype tool, we performed data quality validation on two batches of ML training datasets. The two batches of ML training datasets were of sizes 16,8 MB (ML dataset-1) and 19,8 MB(ML dataset-2). The data validation tests performed were for redundant values (duplicates), features present (new/missing features from baseline), features have right values (Not in Range) and outliers. Summary of these data validation tests is presented in Table \ref{tab:eval_results}. Overall, the two batches of ML training datasets had 620 features in common which were used as a baseline. Huge differences in the number of features were observed from this baseline in both batches. Also several duplicates were detected from the two datasets, while most features had expected values. Given the used ML model, data discrepancy e.g. differences in number of features makes retraining less frequent. Ultimately, new ML models are redeployed rather than improving the current deployed  ML models. Currently, the ML models not frequently redeployed or updated.

\begin{table}[h]
    \centering
  \begin{tabular}{|l|l|l|}
    \toprule
    \textbf{Data validation tests} & \multicolumn{2}{c}{\textbf{Results}}  \\ \cline{2-3}
                    & \textbf{ML dataset-1} & \textbf{ML dataset-2}\\
    \toprule
         New features & 163 & 257 \\ 
         Duplicates & 33.7\% & 22.9\%\\
         Not in Range (Min-Max) & 0.0016\%& 0\%\\
         Outliers & 0.83\% &0.09\% \\
        \toprule
    \end{tabular}
    \caption{Data errors in two batches of ML training datasets}
    \label{tab:eval_results}
\end{table}

Additionally, feedback and discussion sessions were conducted with a total of five practitioners (data scientists, software developer and line manager) within the analytics R\&D organisation on advantages and disadvantages of data validation process and tool. In all sessions, the researcher took notes that were used for further analysis. These qualitative evaluations are presented as benefits and barriers in the results section \ref{sec:findings}.

\subsection{Specifying learning}
Lastly, general lessons from action research are specified based on the evaluations to help decide on how to proceed~\cite{Petersen2014, Staron2020}. Based on the evaluations, lessons learned are presented as best practices for adopting data validation in results section \ref{sec:findings}, and a proposal for a data validation framework (DVF), presented in section \ref{sec:DVF}. 

\section{Results}
\label{sec:findings}
This section discusses experiences of adopting a data validation process and tool for ML projects by data science teams at a large software-intensive company in telecommunication domain. The experiences are shaped in form of \textit{Best Practices, Benefits and Barriers} of adopting data validation in ML projects.
\\
\subsection{Best Practices (RQ1)}
\noindent Best practices of adopting a data validation process and tool for ML projects are classified in three groups: \textit{1) defining data quality tests}, \textit{2) providing actionable feedback}, and \textit{3) treating data errors with similar rigor as code}. 

\subsubsection{\textbf{Defining data quality tests}}
a data validation process for ML projects requires having an overview of the level of data (\textit{feature, dataset, cross-dataset, data stream}) at which certain data quality tests can be applied. Much of our early work was spent on defining the types of data quality tests and how these tests can be performed for the specified ML project.

\textit{Data validation tests at feature and dataset level.} data validation tests are commonly applied at feature and dataset level. Example of data validation tests include checking for \textit{missing values, unique value, range values and outliers}. These encompass majority of what is typically performed by data scientists during data wrangling and feature engineering. 

Typically data validation tools use techniques, such as descriptive statistics to implement declarative tests. In our action research, data scientists wanted to check for redundant values in training dataset wherein a unique entry was determined by using IDs from all sources i.e., Log A, Log B and the external tool. This, together with other checks e.g. outliers, very rare categorical values (Tools hits), all-zero/constant values for all features were implemented in the prototype tool.

\textit{Data validation tests across datasets.}
data validation across two or more batches of datasets constitute another level of data quality tests. Most times new datasets arrive in batches at periodic intervals, such as hourly or daily, and are used to retrain new ML models. Data validation tests between two or more dataset batches aim to detect changes in data distribution and determine \textit{skew}. 

Our data validation tool uses the desired characteristics of data encoded in a data schema to validate new dataset against previous datasets. In our case, we used schema validation to identify new/missing features and out or range values across datasets. Since distance measures are used to determine skew between two datasets, we wanted to use `Jensen–Shannon divergence' measure \footnote{Jensen–Shannon divergence - https://tinyurl.com/y3lc7zxb} similar to TensorFlow data validation but this required the two ML training datasets to have similar dimension. Currently we are exploring measures to detect drift that will not require us to remove instances of the ML training dataset.

\textit{Data validation tests in data streams.}
data validation in real-time streaming data as opposed to batch datasets were qualitatively explored in the study due to the consideration of whether to include the approach and tool before data ingestion phase (i.e., at the cluster or Source dataDB). Typically, streaming processing frameworks and tools, such as Apache Spark in the field study, are used for data conversions in data streams. Data streams imply extremely fast changes in a non-stationary manner. 

Since data becomes instantly available as an event occurs (i.e., has timestamped attribute), for data validation tool, a combination of specific data tests from feature/dataset level and cross-dataset level are used. 

\subsubsection{\textbf{Providing actionable feedback}}
communicating the output results of data validation in terms of warnings and validation report requires a careful design decision of what and how to present to the user. Importantly, data validation tool should include the capability to retrieve failed records and suggest further actions of how to improve data depending on the selected ML algorithms. 

\textit{Data testing report.} 
our data validation tool presents a summarized results of data testing in a textual format. In our data validation prototype tool, if the user performs unique value data testing at feature/dataset the results show the ratio of unique values for each feature contained in the dataset.

\textit{Mitigation strategy.} based on data testing results, the user can perform data cleaning and transformation for records that failed data tests.
Different strategies that take into consideration ML use case and algorithms can be suggested to the user to make data testing report actionable. In our study, missing values are not deleted rather imputed with a fixed value. For new features that are only available in subsequent dataset but not in previous dataset, data scientist gets a warning and suggestion of how to improve the data schema. When there is a huge deviation in data distribution across datasets, a suggestion is to further explore data sources and confirm the anomalies in the collected data. 

\subsubsection{\textbf{Treating data errors with similar rigor as code}}
similar to software bugs, data errors should be documented, tracked and resolved. Like software unit tests that try to test atomic components in codebase, data validation tests allow designers to quantify the performance of ML models adhering to some specific properties found in ML training datasets. Therefore, structuring data validation tests around the properties of data that the ML model expects to acquire serves as one such approach. In our study, collaborative work, which is important in ML projects, resulted in several instances of knowledge transfer between data scientists and software engineers on best practice from the different disciplines.

\subsection{Benefits (RQ2)}
The benefits of adopting data validation process and tool for ML projects include: \textit{1) minimization of manual effort in data preparation; 2) early identification of data errors; 3) a testing approach to ML enabled software systems}. 

\subsubsection{Minimization of manual effort in data preparation} among the main benefits for data validation tool in ML pipeline are minimized manual efforts when preparing data for an ML model (re)training. In addition, as team members are typically involved in several projects, automated data validation allows scaling data validation practice across the different projects. Particularly in our study, it reduced the copy-paste of regular data cleaning and processing steps often making data scientists to work more efficient.

\begin{quote}
    \textit{Automatic data validation reduces the manual effort of preparing data before modelling. It is a good auxiliary to ease data scientists daily work. It defines the quality of data and brings the potential to find noise in the early stage of featuring engineering. In a streaming solution, automatic data validation could help with deriving invalid data before processing. (Data scientist 1)} 
\end{quote}

\begin{quote}
    \textit{Even the smallest of data validation checks, like dataset size, should be automated because you do not want to rewrite the code (or documentation for its implementation) every time during ML  model training (Data scientist 2)}
\end{quote}

\subsubsection{Early identification of data errors} data validation provides the capability to monitor what is happening with ML training dataset continuously. For example, in our study teams are able to identify and detect when incoming new ML training dataset breaks any rules especially since different joins have been used to create ML training datasets. However, in our study adopting data validation after creating ML training dataset to for example identify duplicates required the data scientist to further investigate their origins. Monitoring data errors at a stream is preferred particularly because data may be useful to other teams (data reuse for multiple purposes). Furthermore, when introducing a new data source, data validation can be performed to help with understanding and troubleshooting data quality issues. 

\begin{quote}
    \textit{If we run data validation in production, we can receive alerts if there are any major changes in the distribution of the data...if the incoming datasets breaks any rules (such as null values appearing where it should not). (Data scientist 3)} 
\end{quote}

\begin{quote}
    \textit{``We can run the tool to decide if a new data source is ready for production (e.g., we receive data from a new customer that is missing some particular values) or to decide if any data source starts to have missing mandatory files." (Data scientist 2)}
\end{quote}

\subsubsection{A testing approach for ML enabled software systems}
data validation can be used to facilitate regular testing for ML-enabled software system, for example in all ML training dataset creation jobs to identify data errors but also bugs in code changes. In the studied context, data scientists are typically expected to write unit tests for their code, including the code used to check data issues. For the ML component it is not enough to write simple unit tests with sampled datasets but rather requires testing an entire ML training dataset. The data validation tool, as a well-tested system, helps to standardize the tests across entire datasets and is usable for different ML projects. 

\begin{quote}
    \textit{There is a shift in the way of writing tests from defining input data and expected output to defining properties of the input and expected output of data. This allows developers of ML system to use the same properties and test for any input data and can be use at any scope of testing e.g., unit, integration, system and production tests (software developer)}
\end{quote}

\subsection{Barriers (RQ3)}
The barriers of adopting data validation include \textit{1) limited flexibility of data validation tool e.g. in terms of ease of adding new tests, and 2) limited support for the existing technology stack of ML system development process while also ensuring low learning curve. }

\subsubsection{Limited flexibility of data validation tool e.g. in terms of ease of adding new tests}
the scope of data validation tool like the number of tests are covered and ease of adding new tests. According to the data scientists, it is important to involve several stakeholders from other teams to help in identifying additional data validation tests that can be implemented by the tool. This will also increase the use of the data validation process and tools across the different teams within the organization. Different teams with a stake on the data should be able to define and implement new data checks to the tool while ensuring good usability in terms of appropriate mechanism to alert users and suggest ways to improve data quality.

\begin{quote}
    \textit{``If it is not easy to add new checks to the tool or if it is not easy to maintain, the data scientists/engineers might decide to go back to ad-hoc solutions." (Data scientist 2)}
\end{quote}

\subsubsection{Limited support for the existing technology stack of ML component while also ensuring low learning curve} In general, data validation tool needs to support existing technology stack used to develop the ML component for performance and maintenance. If highly decoupled, the tool needs to ensure low learning curve on how to use  and extend the tool. With regards to technology stack support, prototype tool uses libraries familiar and extensively used by the data scientists.

\begin{quote}
    \textit{``If the tool has low performance (cannot handle larger datasets) and it is a crucial step to pass before delivering datasets, it becomes a bottleneck in the infrastructure " (Software developer)}
\end{quote}

\section{Data validation Framework (DVF)}
\label{sec:DVF}
\begin{figure*}[ht!]
    \centering
    \includegraphics[scale=0.7]{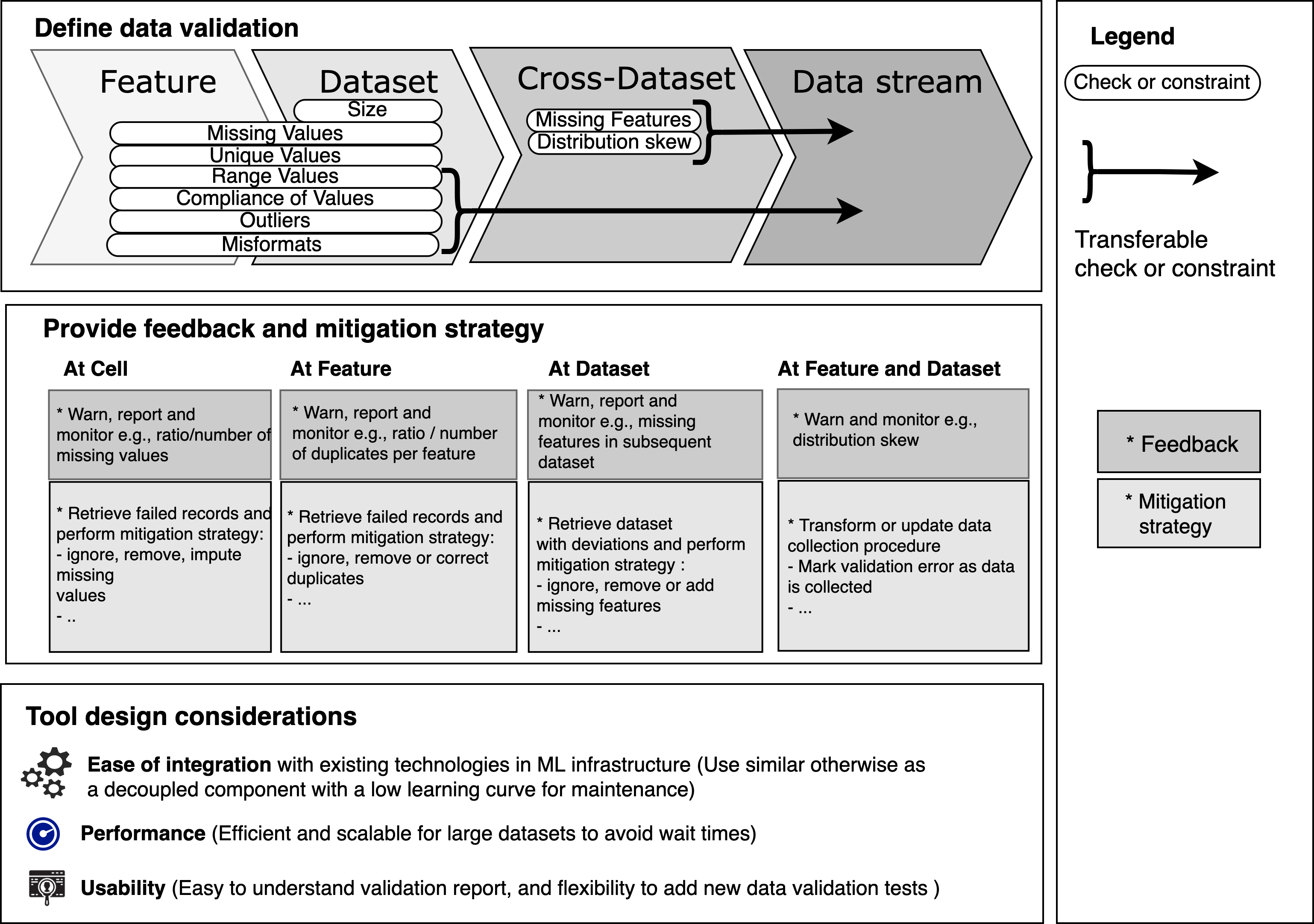}
    \caption{A data validation framework (DVF) exemplified with checks from studied ML use case}
    \label{fig:validation-figure}
\end{figure*}

Based on the experiences, we propose a data validation framework (DVF) shown in Figure \ref{fig:validation-figure} that systematizes the adoption of data validation in ML projects. We group important aspects in the DVF into: \textit{A) validation process},  \textit{B) validation artefacts}, \textit{C) data validation types, \textit{D)  data validation tool setup, and \textit{E) feedback and mitigation strategy}.}}

\subsection{\textbf{Validation process}} The data validation process is closely related to the process of collecting and preparing data, and to the training process as also shown in \ref{fig:data-pipeline}. The process includes an \textit{initial stage} and \textit{operational stage}.

\textit{The initial stage} is a part of the initial dataset setup: identification of input data, selection of features and creation of new features, and other input parameters required for ML training process. The data validation part includes specification of (and implementation of needed) data validation tests.

An ML project sets functional requirements that form as bases for data selection, and quality requirements on this data in order to accurately train ML models. Some data quality requirements are determined iteratively during experimentation of ML models prior to deploying them to production. The identified data quality requirements are used to derive data validation tests, which test specific data properties. Data validation tests are defined by taking into consideration the ML model that hold certain assumption on data e.g. data should not contain missing values. Also, while different data formats will require different mechanisms to create ML training dataset, they will also employ different techniques to facilitate different data validation tests. For example, when validating image data additional techniques that measure inter-rater agreement in annotated labels are used. To ensure a more general data validation process and tool that is applicable to various ML projects, more stakeholders need to be involved in defining data quality requirements that extend beyond a specific ML project. In our case study, the initial stage includes the decisions at which stage of the collection data the data validation tests will be provided, as shown in Figure~\ref{fig:data-pipeline}, and which data validation tests will be performed. 

\textit{The operational stage} includes a setup of execution of the data validation tests through the initial training and the run-time stage. These setups can be different: data validation tests can be performed a) only once if the input data are not changing, b) re-run for every new project in which different system requirements can cause different validation requirements on data features and quality metrics of ML, c) periodically, or on demand, when input data has been changed. For each new validation, the data validation tool might need to be updated, which should be analysed as a part of the process. In our case study, the  data validation tests are performed periodically, as the input data are collected continuously in a Source dataDB, then asynchronously created as periodical logs and delivered in batches to a common dataset. That requires a periodical update of the training set and its validation.

\subsection{\textbf{Validation artefacts}} The basic and obvious data validation artifacts are the dataset(s) and the features in the dataset(s), or features selected for particular goals as part of a project. However, complex systems often use different datasets and input data is collected in different way. In some cases data are collected once, and in other cases data are collected continuously in a form of a stream, or as batches. These artefacts require their data validation, but also comparative validation between them. Identification of the artefacts are followed by the identification and processing of data validation.

An ML training dataset contains N-dimensional feature representation of data. At the lowest level, the elicited data quality requirements can be tested at individual feature contained in ML training dataset. However, often time data quality tests are used to verify data properties across the entire ML training dataset. For periodic updates of ML training dataset, an already verified batch of dataset can be used instead to verify subsequent batches to understand how the different data properties have changed and if they are expected or not. Data quality tests at data streams verify individual data events as they occur thereby influencing data collection process. In our case study, the validation artifacts include a) dataset, b) feature, and c) cross-dataset (batches). However, data scientists questioned the possibility to perform data validation tests at the source, e.g., data stream.

\subsection{\textbf{Data validation types}}Data validation types defines what is of interest to validate in a project or for a given dataset. In practice, there are a number of data validation types. Standard types of validation are related to size (e.g. size of dataset, number of features), data values (e.g. unique values, range values, outliners), quality of data (missing values, value accuracy, value reliability, time-related value, missformats), cross-artifact data validation (missing features, skew). Different types are applied for different artefacts). In our case study, a subset of the predefined types has been used as seen in Figure \ref{fig:validation-figure}.

\subsection{\textbf{Data validation tool setup}} For the identified validation types, artifacts and process, the tools that perform the validation need to be setup. As discussed in Section \ref{sec:background}, there are several tools available for data validation with their advantages and limitations, such as usability, integration possibilities, performance, and functionality. In a case where the constraints are larger than benefits, a validation tool set can be built locally.

In our case, the judgment was that an introduction of existing data validation tools required too much efforts in terms of high learning curve of the new technology for maintenance, including also the need to support dependencies not currently in the ML infrastructure. As a result, a validation prototype tool was implemented, as described in Section \ref{sec:methodology}. Since this was the first project using DVF, and because of its benefits several improvements related to usability and performance of the data validation prototype identified are being worked on in order to be reused and adapt to the new requirements of other ML projects.

\subsection{\textbf{Feedback and mitigation strategy}}After running data quality tests, data validation process and tool is only useful if it the feedback is actionable, i.e it provides verification report containing different warnings and alerts as well as a way to handle the identified data anomalies. A mitigation strategy to handle data anomalies will require data retrieval and transformation at different data granularity depending on the level at which data quality tests are performed. For example, handling data anomalies at dataset level will cover features while at data stream influence cover entire dataset. Figure \ref{fig:validation-figure} shows a set of feedbacks and mitigation strategies in relation to particular validation artefacts, and validation process.

\section{Discussion}
\label{sec:discussion}
Data validation tools are increasingly becoming incorporated in ML pipelines to ensure that not only data errors are caught early but also improve the ML model understanding and debugging~\cite{breck2019Google}. In this study we explored the adoption of data validation process and tool using action research in R\&D analytics organisation at a large software-intensive company within telecommunication domain.

The main results of the study show that it is crucial to have an understanding of the different and useful data validation tests to be implemented at different levels of data (feature, dataset, cross-dataset, datastream) when adopting data validation process and tool in ML projects. In addition, data validation tools needs to have the capability to not only warn users of data anomalies but give actionable feedback by suggesting and support appropriate actions to take in order improve data quality. Our proposed DVF framework for adopting a data validation process and tool in ML projects provides an approach for defining data validation tests at different levels of data and suggestions for improving data quality. Among the design principles of Tensorflow data validation tools~\cite{Baylor2017tfx} that affected the logic to detect and present data anomalies included the ability for users to at a glance understand data quality checks over the data covered and detected data errors should have a description of how to debug and fix the data.

Although data cleaning is an extensively studied problem~\cite{Chu2016DataCleaning}, there is limited support for data cleaning and transformation functionality in existing data validation tools. We observed that data cleaning is often decoupled from the actual data validation. Previous studies reported that user engagement in data cleaning tasks, such as repairing detected data errors and using user-feedback in improving data quality checks, still remains to be explored~\cite{Chu2016DataCleaning}. Researchers~\cite{Krishnan2016ActiveClean} proposed a framework, ActiveClean, to help control the impact of data cleaning in downstream ML models. The approach in ActiveClean framework focuses on incremental cleaning of training dataset making data cleaning step aware of subsequent step (e.g., model training) to ensure good performance of ML models. In each data cleaning increment, sampling techniques are used to select and visualize a sample of likely dirty records from training dataset through diagnose interface. At the cleaning interface, the user has the option of removing dirty records, selecting pre-defined list of cleaning functions or apply a custom cleaning operation. ActiveClean is useful during ML model experimentation where the characteristics of ideal ML training dataset are being explored.

While it is difficult to generally describe how data quality should be measured, Ehrlinger et al.,~\cite{Ehrlinger2019Siemens} took a practical approach and focused on detecting data errors that were relevant to ML data of their industrial partner (Siemens). The data validation checks included: missing values, outliers, and values outside a given range~\cite{Ehrlinger2019Siemens}. The proposed data quality library for monitoring the quality of data employed for ML application require extension of other measures, such as detection of duplicates and expansion from batch-based to stream-based execution~\cite{Ehrlinger2019Siemens}.

The data validation process and tools at datastreams is an open research area since most tools have focused on batch data processing~\cite{Swami2020LinkedIn}. However, data validation at data streams is feasible and requires careful design considerations for monitoring data quality issues and mitigation strategies during data conversion because they affect subsequent stages of ML development process. One common approach to detect anomalies in datastreams is by using a ML approach where an ML model is created based on historical data representing normal behaviour. Pizonka et al~\cite{pizonka2018domain} proposed a complementary approach in which validation rules are derived from pre-defined domain models and are being interpreted in a stream processing framework at run-time. The authors~\cite{pizonka2018domain} presented a prototype implementation of automated data validation at data stream for IoT applications that combines domain model with stream processing system, thereby allowing both syntactic and semantic validation. The domain model provides high-level description of functionality and characteristics of IoT devices that contain validation related information, such as allowed range values of humidity for micro-sensor boards~\cite{pizonka2018domain}. Although their implementation showed that some validation errors can go undetected, data validation approach was seen feasible at data stream level. However, how data cleaning approaches will work on distributed streams of data remains an area for future research~\cite{Chu2016DataCleaning}.

The data validation process and tools for time-series data, other data types, such as images~\cite{Neila2019Image} and specific class of AI problems, such as natural language processing (NLP) will require different checks currently not specified in the DVF framework. For identifying corrupted images in input data,~\cite{Neila2019Image} proposed a data validation method that is based on ML. The method focuses on learning the appearance of images from previous training dataset used to train a deep neural network, giving the ability to identify an input image that deviates from the training distribution~\cite{Neila2019Image}. In addition to the ability of detecting images not belonging to the training distribution, other image quality attributes for consideration include distortions caused by image compression, image sharpness or blurring, entropy from image representations based on histograms. The data validation process and tools for image data is much broader and requires much detailed research.

The data validation process and tools consume huge amounts of engineering resources, including mainteance~\cite{Swami2020LinkedIn}. Increasingly there is huge growth in commoditized technology that can be challenging for practitioners to decide from the myriad of data validation tools~\cite{ehrlinger2019survey,breck2019Google, Baylor2017tfx,hynes2017Google,Schelter2018Amazon,Swami2020LinkedIn}. As reported in the results of our study, ease of integration with existing technology stacks into ML development process and infrastructure at the organization is crucial for their adoption. This latter was also observed in other studies~\cite{Swami2020LinkedIn}.

\subsection{Validity threats and limitations of the study}
A strength of our study is the active involvement of industrial practitioners. However, several factors present some potential threat to the validity of the study.
\begin{itemize}
    \item Construct validity is related to research design and the measurement instruments that measure the effects in the study~\cite{Staron2020}. Access and collaborative work with practitioners helped with the investigations, but time constraints limited exhaustive evaluations e.g., in other teams.
    \item Internal validity concerns scrutinizing the operation of the action taking~\cite{Staron2020}. A main internal validity related to the maturity of the analytics R\&D organisation which was evolving at fast pace both in personnel turnover but also introduction of new ways of working. During this period, developers were interested and involved in several process improvement initiatives, including infrastructure migrations to the cloud.
    \item Regarding generalizability of the results, more empirical studies need to be conducted to identify additional best practices as well as validate and extend DVF framework.
    
\end{itemize}
\section{Conclusion and future work}
\label{sec:conclusion}
Best practices, benefits and barriers of adopting a data validation process and tool in ML projects at a large-software intensive organization within telecommunication industry were investigated by using action research. Prior studies identified have focused on the implementation details of data validation tools and lack best practices for adoption especially by development teams in the early stages of operationalizing ML models. The main contribution of the study is a data validation framework (DVF) that provides an approach for systematizing the adoption of data validation process and tool for ML projects taking incorporating best practices identified by this study. The DVF considers the definition of data validation checks at multiple levels of data and provision of feedback and strategy for the next step of improving data quality. In future research we aim to focus on extending the proposed DFV framework to other ML use cases, including NLP-based AI problems. In addition also identify design principles for data validation in data streams as with others e.g. \cite{Swami2020LinkedIn}.


\bibliographystyle{IEEEtran}

\small
\bibliography{main}

\end{document}